\documentclass[11pt,letterpaper,singlespacing, abstract=on]{scrartcl}
\usepackage[utf8]{inputenc}
\usepackage[usenames,dvipsnames]{xcolor}
\usepackage{amsmath,amssymb, amsfonts, amsthm, mathtools} 
\usepackage{tikz}
\usepackage{subfigure}
\usepackage{microtype}
\usepackage{paralist}
\usepackage{stmaryrd}
\usepackage{fancyhdr}
\usepackage{tabularx}
\usepackage[margin=1in]{geometry}
\usepackage{setspace}
\usepackage{array}
\usepackage{titlesec}
\usepackage{parskip}
\usepackage[pdfpagelabels=true]{hyperref}
\usepackage{listings}
\usepackage[T1]{fontenc}
\usepackage{etoolbox}
\usepackage{standalone}

\usetikzlibrary{decorations.pathreplacing}

\definecolor{darkgreen}{rgb}{0,0.6,0}

\def\Nscr{\mathcal{N}}

\def\Lscr{\mathcal{L}}

\def\cupp{\stackrel{.}{\cup}}
\newcommand{\sfrac}[2]{{\textstyle\frac{#1}{#2}}}

\renewcommand{\epsilon}{\varepsilon}
\renewcommand{\phi}{\varphi}

\newtheoremstyle{mytheorem}{0.5cm}{0.5cm}{\itshape}{}
   {\bfseries}{}{\newline }{#1\, #2\, #3}
\theoremstyle{mytheorem}

\newtheoremstyle{note}{0.5cm}{0.5cm}{\upshape}{}
   {\bfseries}{}{\newline}{#1\, #2\, #3}
\theoremstyle{note}

\theoremstyle{mytheorem}
\newtheorem{theorem}{Theorem}

\theoremstyle{mytheorem}

\theoremstyle{mycorollary}

\theoremstyle{mytheorem}
\newtheorem{lemma}[theorem]{Lemma}

\theoremstyle{mytheorem}

\theoremstyle{note}

\theoremstyle{note}

\def\cupp{\stackrel{.}{\cup}}
\def\prove{\noindent \hbox{\textbf{Proof:}}\quad}

\def\endproof{\hspace*{\fill} {\boldmath $\Box$} \par \vskip0.5em}

\begin{document}

\title{\boldmath An improved upper bound on the integrality ratio for the $s$-$t$-path TSP}
\author{Vera Traub \and Jens Vygen}
\date{\small Research Institute for Discrete Mathematics and Hausdorff Center for Mathematics,\\
University of Bonn \\
\texttt{\{traub,vygen\}@or.uni-bonn.de}}

\begingroup
\makeatletter
\let\@fnsymbol\@arabic
\maketitle
\endgroup

\begin{abstract}
We give an improved analysis of the best-of-many Christofides algorithm with lonely edge deletion, 
which was proposed by Seb\H{o} and van Zuylen \cite{SebvZ16}. 
This implies an improved upper bound on the integrality ratio of the standard LP relaxation
for the $s$-$t$-path TSP.
\end{abstract}

\section{Introduction}

A major open problem in the study of the traveling salesman problem is to determine the integrality ratio of 
the standard LP relaxation. The same question can be asked for the variant in which start and end of the tour are 
given and distinct. 
For this LP (see \eqref{eq:subtour_lp_with_degree} below) the conjectured integrality ratio is $\frac{3}{2}$,
which is asymptotically attained by simple examples.
Better and better upper bounds have been shown \cite{AnKS15, Seb13, Vyg16, GotV16, SebvZ16}.
The previously best-known upper bound by Seb\H{o} and van Zuylen \cite{SebvZ16} is $\frac{3}{2}+\frac{1}{34}>1.5294$.
We improve the analysis of their algorithm and show that the integrality ratio is smaller than $1.5284$.

Even better approximation algorithms, with ratios $\frac{3}{2}+\epsilon$ \cite{TraV18} and $\frac{3}{2}$ \cite{Zen18}, have been found recently,
but these do not imply an upper bound on the integrality ratio of the LP.

In the $s$-$t$-path TSP we are given a finite metric space $(V,c)$ and vertices $s,t\in V$ with $s\ne t$. 
The task is to compute a path $(V,H)$ with endpoints $s$ and $t$
that contains all elements of $V$ and minimizes $c(H) \coloneqq \sum_{\{v,w\}\in H} c(v,w)$.
An equivalent formulation asks for a minimum-cost multi-subset $H$ of $E= {V \choose 2}$ such that the graph
$(V,H)$ is connected and $s$ and $t$ are its only odd-degree vertices. 
(Such a graph contains an Eulerian $s$-$t$-walk and we can shortcut whenever we revisit a vertex.)

Christofides' algorithm, adapted to the $s$-$t$-path TSP by Hoogeveen \cite{Hoo91}, computes a minimum-cost spanning tree $(V,S)$
and adds a minimum-cost matching on the set $T =\{v \in V : |S\cap \delta(v)| \text{ odd }\}\triangle \{s,t\}$ 
of vertices whose degree has the wrong parity. 
Adding such a matching is called parity correction.
This algorithm yields a tour of cost at most $\frac{5}{3}$ times the LP value.

\section{Best-of-many Christofides with lonely edge deletion}

An, Kleinberg and Shmoys \cite{AnKS15} proposed and analyzed the best-of-many Christofides algorithm. 
It starts by computing an optimum solution $x^*$ to the LP
 \begin{equation} \label{eq:subtour_lp_with_degree}
 \begin{aligned}
 &\min c(x) \\
 &\text{s.t.} & x(\delta(U)) &\geq 2  & & \text{for } \emptyset \subset U \subseteq V\setminus\{s,t\},\\
 & & x(\delta(U)) &\geq 1  & & \text{for } \{s\} \subseteq U \subseteq V\setminus\{t\}, \\
 & & x(\delta(v)) &= 2  & & \text{for } v\in V\setminus\{s,t\}, \\
 & & x(\delta(v)) &= 1  & & \text{for } v\in \{s,t\}, \\
 & & x(e) &\geq 0 & & \text{for } e\in E.
\end{aligned}\hspace{-4mm}
\end{equation}
Here and in the following we abbreviate $x(F) = \sum_{e\in F} x(e)$ and $c(x) \coloneqq \sum_{e\in E} c(e)x(e)$.
We write $x^*$ as a convex combination of incidence vectors of spanning trees, i.e. 
$x^* = \sum_{j=1}^k p_{j} \chi^{S_j}$ for spanning trees $(V,S_1), \dots, (V,S_k)$ and nonnegative coefficients 
$p_{1}, \dots, p_{k}$ with $\sum _{j=1}^k p_{j} =1$. 
Then parity correction as in Christofides' algorithm is applied to each of the $k$ spanning trees; finally 
the best of the resulting $s$-$t$-tours is selected. 
A key observation of \cite{AnKS15}, used in all subsequent works, was that the set 
$\Nscr \coloneqq \{ \delta(U) : \{s\} \subseteq U \subseteq V\setminus\{t\},  x^*(\delta(U)) < 2\}$
of \emph{narrow} is induced by a chain.
The analysis of the best-of-many Christofides algorithm was improved by Seb\H{o} \cite{Seb13}.
The algorithm can be further improved by using a convex combination with certain properties \cite{Vyg16,GotV16}.
In particular, Gottschalk and Vygen \cite{GotV16} showed:
\begin{theorem}\label{thm:gao_tree_decomposition}
Let $x^*$ be an optimum solution to the LP \eqref{eq:subtour_lp_with_degree} and $\Nscr$ the set of narrow cuts.
Then there exist
spanning trees $(V,S_1), \dots, (V,S_k)$ and nonnegative coefficients $p_{1}, \dots, p_{k}$ with $\sum _{j=1}^k p_{j} =1$
such that $x^* = \sum_{j=1}^k p_{j} \chi^{S_j}$ and for every $C\in\Nscr$ there exists an $r\in\{1,\ldots,k\}$ with 
$\sum_{j=1}^r p_{j} = 2-x^*(C)$ and $|C\cap S_j|=1$ for all $j=1,\ldots,r$.
\end{theorem}

Schalekamp et al.\ \cite{SchXX18} found a simpler proof of this theorem.
We will work with such a convex combination henceforth.

Seb\H{o} and van Zuylen \cite{SebvZ16} had the brilliant idea to delete some of the edges in each spanning tree and do parity correction
on the resulting forest. 
This can save cost because parity correction will often reconnect the connected components of the forest anyway.
%Different from Section \ref{sec:32}, but consistent with \cite{SebvZ16}, 
Call an edge $e$ and a cut $C\in\Nscr$ \emph{lonely} in tree $S_j$ 
if $\{e\}=C\cap S_j$ and $\sum_{i=1}^j p_{i} \le 2-x^*(C)$.
Then we also say that $e$ is lonely at $C$.
We denote the lonely cuts in $S_j$ by $\Lscr(S_j)$.
Let $F_j$ be the edge set of the forest that results from $S_j$ by deleting its lonely edges.
The algorithm by Seb\H{o} and van Zuylen \cite{SebvZ16} does parity correction on each forest $(V,F_j)$.
Let $T_j\coloneqq\{v\in V:|F_j\cap \delta(v)| \text{ odd}\}\triangle\{s,t\}$ 
denote the set of vertices whose degree in $F_j$ has the wrong parity.

Instead of adding a matching on $T_j$, we can add an arbitrary $T_j$-join $J$ (an edge set such that $T_j$ is the set odd-degree vertices 
of $(V,J)$). Although this is equivalent, it will come handy.
Every $T_j$-join $J$ must contain an edge (in fact, an odd number of edges) in every lonely cut of $S_j$ (because they are all $T_j$-cuts,
i.e.\ cuts $\delta(U)$ for a vertex set $U$ with $|U\cap T_j|$ odd).
% \red{more explaination?}
However, this does not imply that $F_j\cupp J$ is connected, because an edge of $J$ can belong to several lonely cuts of $S_j$.
In this case we can, for all but one of these cuts, add two copies of the lonely edge of $S_j$ in this cut
(to ensure connectivity without changing parities).
%(which belongs to $S_j\setminus F_j$).

If we choose a $T_j$-join $J$ for parity correction, we will pay a total of at most $\sum_{e\in J} c^j(e)$, where
$$c^j(e) \ \coloneqq \ c(e) \ +  \sum_{C\in\Lscr(S_j),\, e\in C} 2c(C\cap S_j) \ - \ \max\left\{0,\, \max_{C\in\Lscr(S_j),\, e\in C} 2c(C\cap S_j) \right\};$$
here the second and third terms account for the reconnection cost.

We now describe formally the best-of-many Christofides algorithm with lonely edge deletion due to Seb\H{o} and van Zuylen.
This is the algorithm that we will analyze.
\begin{enumerate}
\item Compute an optimum solution $x^*$ to the LP \eqref{eq:subtour_lp_with_degree}.
 \item Compute $x^* = \sum_{j=1}^k p_{j} \chi^{S_j}$ as in Theorem \ref{thm:gao_tree_decomposition}
\item Do the following for each $j=1,\ldots,k$:
\begin{enumerate}
\item  Compute a $T_j$-join $J_j$ with minimum $c^j$-cost.
 \item Compute a minimum $c$-cost subset $R_j \subseteq S_j \setminus F_j$
       of the lonely edges such that $F_j \cupp J_j \cupp R_j$ is connected. 
 \item Find an Eulerian $s$-$t$-walk in $H_j \coloneqq F_j \cupp J_j \cupp R_j\cupp R_j$ and shortcut whenever a vertex is visited more than once.
\end{enumerate}
\item Return the cheapest of these $k$ tours.
\end{enumerate}

We remark that Seb\H{o} and van Zuylen \cite{SebvZ16}
also consider the result of the normal best-of-many Christofides algorithm and output the better of the solutions, 
but this is not necessary as our analysis will reveal. 

\section{Outline of the new analysis}

By definition of $c^j$, the cost of the tour $H_j$ is at most $c(F_j) + c^j(J_j)$.
The cost of the $T_j$-join $J_j$ is the minimum cost of a vector $y$ in the $T_j$-join polyhedron \cite{EdmJ73}
\begin{equation}\label{eq:Tjoinpolyhedron}
\bigl\{y\in\mathbb{R}^E_{\ge 0} : y(\delta(U))\ge 1 \text{ for $ U\subset V \text{ with } |U\cap T_j|$ odd} \bigr\}.
\end{equation}
We call a vector $y$ in \eqref{eq:Tjoinpolyhedron} a parity correction vector.
Note that every parity correction vector yields an upper bound on the cost of $J_j$.
A first attempt to design a parity correction vector could be the vector $\beta x^* + (1-2\beta) \chi^{S_j}$ for some $0\le \beta \le \frac{1}{2}$.
This vector has value at least one on all cuts except the narrow cuts.
The narrow cuts can be repaired by adding fractions of incidence vectors of lonely edges (not necessarily from the same tree).
We will pay all this and the reconnection cost by what we gain by deleting the lonely edges.
Then our total cost is 
\begin{equation}\label{eq:uniform_weights}
 \min_{j=1}^k c(H_j) \le \sum_{j=1}^k p_j c(H_j) \le \sum_{j=1}^k p_j \bigl(c(S_j) + \beta c(x^*) + (1-2\beta) c(S_j)\bigr) = (2-\beta) c(x^*).
\end{equation}
Hence we would like to choose $\beta$ as large as possible.
Unfortunately, for $\beta = \frac{1}{2}$ we need too much from the lonely edges. 
By reducing $\beta$, we can increase the value of $\beta x^* + (1-2\beta) \chi^{S_j}$ on the narrow cuts and thus decrease the 
required amount of lonely edges.
Choosing $\beta = \frac{8}{17}$ is sufficient and this is essentially what Seb\H{o} and van Zuylen did.

%The cost of the tour arising from the first tree $S_1$ is at most $c(S_1) + \frac{1}{2}c(x^*)$.
In our parity correction vector for a forest $F_j$ we will use lonely edges of $S_j$ and of earlier trees.
If we increase $\beta$ for the early trees and decrease $\beta$ for the late trees, we need more from the lonely edges 
in the early trees, but less in the late trees. 
This will improve our bound if the late trees are cheaper (and this is indeed true in the worst case).

The algorithm computes $k$ tours $H_1,\ldots,H_k$. 
All previous analyses, like \eqref{eq:uniform_weights}, computed an upper bound on $\sum_{j=1}^k p_j c(H_j)$. 
Instead, we will compute a weighted average with different weights, giving a higher weight to tours resulting from early trees.

We choose $\beta_j$ and weights $q_j >0$ with $\sum_{j} q_j = 1$ and such that $q_j \cdot (2-2\beta_j) = M \cdot p_j$ for some constant $M>0$.
Such a choice allows to bound the cost of our tour against the LP value $c(x^*)$:
\[\sum_{j=1}^{k} q_j \cdot c(H_j) \le \sum_{j=1}^{k} q_j \left(\beta_j c(x^*)
 + (2-2\beta_j) c(S_j)\right) = \left(\sum_{j=1}^{k} q_j \beta_j +  M\right) c(x^*). \]
Intuitively, choosing $\frac{q_j}{p_j}$ (and thus $\beta_j$) larger for the early trees is good, because for the early trees
we delete more lonely edges (cf.\ Theorem \ref{thm:gao_tree_decomposition}).
This allows us to choose the average value of $\beta$ larger and thus improves our upper bound.

We first analyze the cost of a tour resulting from a single tree $S_j$.
Later, we will take a weighted average.

\section{Analyzing one tree}

Let $j\in\{1,\ldots,k\}$.
To bound the cost of parity correction of the forest $F_j$, we follow Wolsey's approach \cite{Wol80} and 
use a vector in the $T_j$-join polyhedron \eqref{eq:Tjoinpolyhedron}.

Let $0 \le \beta \le \frac{1}{2}$ and $\alpha \coloneqq 1- 2\beta \ge 0$. 
Moreover, for $C\in\Nscr$, let $v^C\in\mathbb{R}_{\ge 0}^{E}$ be a vector with $v^C(C) = 1$ and 
$v^C(e) = 0$ unless $e$ is lonely at $C$ in some tree (not necesarily in $S_j$).
We will choose $v^C$ later. 
We define 
%For every $\sigma$ with $\sum_{i\in [j-1]} p_i < \sigma \le \sum_{i\in [j]} p_i$ (in other words, for every $\sigma$ with $j=j(\sigma)$), let
\begin{equation*}
y^j_{\beta} \ \coloneqq \beta x^* + \alpha \chi^{S_j} 
+\! \displaystyle\sum_{C\in\Lscr(S_j)} \beta (2-x^*(C)) \chi^{S_j\cap C} 
%&+&\!\!\! 
+\! \displaystyle\sum_{C\in\Nscr\setminus\Lscr(S_j)}\!\! \max\left\{0,\, \beta (2-x^*(C)) - \alpha \big. \right\} v^C.
\end{equation*}
The first sum is the contribution from lonely edges of the tree $S_j$ itself, in oder to repair the lonely cuts of $S_j$.
The second sum is the contribution from lonely edges of earlier trees, in order to repair the other narrow cuts.

%Since $0 \le \beta \le \frac{1}{2}$ and $\alpha = 1- 2\beta$, we have 
%$1-2\beta \ge 0$ and 
Obviously, $y^j_{\beta}$ is a nonnegative vector.
%Similarly to a lemma of An, Kleinberg and Shmoys \cite{AnKS15}, 
We show that $y^j_{\beta}$ is a parity correction vector, i.e.\ a vector in \eqref{eq:Tjoinpolyhedron}.
%can be used to bound the cost of parity correction for $F_j$: 

\begin{lemma}
For every $T_j$-cut $C$ we have $y^j_{\beta}(C)\ge 1$.
\end{lemma}

\prove
Let $C=\delta(U)$ be a $T_j$-cut. 
Since $|\{ v\in U : |F_j\cap\delta(v)| \text{ odd}\}|$ is odd if and only if 
$|F_j\cap \delta(U)|$ is odd, we conclude that
$|U\cap \{s,t\}|+ |F_j\cap C|$ is odd.
We now distinguish several cases. 
\\[2mm]
\underline{Case 1:} $|U\cap  \{s,t\}|$ is odd (i.e., $C$ is an $s$-$t$-cut). \\[1mm]
Then $|F_j\cap C|$ is even. We now consider two subcases.
% Moreover, the intersection of an $s$-$t$-path and an $s$-$t$-cut is odd, so  $|I_j\cap C|$ is odd.
\\[1mm]
\underline{Case 1a:} $C\in\Lscr(S_j)$. \\[1mm]
Then $y^j_{\beta}(C)\ge \beta x^*(C)+\alpha+ \beta(2-x^*(C)) =  \alpha+2\beta = 1$.
\\[1mm]
\underline{Case 1b:} $C\notin\Lscr(S_j)$. \\[1mm]
Since $|F_j\cap C|$ is even, we have $|S_j\cap C| \ge |F_j\cap C| \ge 2$ or $|F_j\cap C| = 0$.
Since $C\notin\Lscr(S_j)$, if $F_j\cap C$ is empty, the cut $C$ must contain at least two edges that are lonely in $S_j$.
So we have also in this case $|S_j\cap C| \ge 2$.
%This shows that either $|I_j\cap C|\ge 3$ or ($|I_j\cap C|=1$ and $|(S_j\setminus I_j)\cap C|\ge 1$).
Thus $y^j_{\beta}(C)\ge \beta x^*(C) +2\alpha + \max\{0,\beta(2-x^*(C))-\alpha\}$
(note that the last term is zero if $C\notin\Nscr$). 
We conclude 
$y^j_{\beta}(C)\ge \beta x^*(C) +2\alpha  + \beta(2-x^*(C))-\alpha
= \alpha+2\beta = 1$.
\\[2mm]
\underline{Case 2:} $|U\cap  \{s,t\}|$ is even. \\[1mm]
Then  $x^*(C)\ge 2$.
% If $|I_j\cap C|=0$, then $|(S_j\setminus I_j)\cap C|\ge 1$, so we have $|I_j\cap C|\ge 2$ or $|(S_j\setminus I_j)\cap C|\ge 1$.
Hence, $y^j_{\beta}(C)\ge \beta x^*(C)+ \alpha \ge 2 \beta + \alpha = 1$.
\endproof

Moreover,we have $y^j_{\beta} \ge 0$. Thus $y^j_{\beta}$ is contained in the $T_j$-join polyhedron \eqref{eq:Tjoinpolyhedron}, and so
$\min \{c^j(J) : J\text{ a } T_j\text{-join} \} \le c^j(y^j_{\beta})$.

A key observation of Seb\H{o} and van Zuylen \cite{SebvZ16} was that the need for reconnection is unlikely.
Only bad edges can result in reconnection, where an edge is called \emph{bad} (for $S_j$) if it is contained in more than one lonely cut.
The edges in $S_j$ are never bad for $S_j$, nor are the lonely edges of trees that come earlier in the list $S_1, \dots, S_r$.
Therefore, an edge $e$ with $v^C(e) > 0$ for some $C\in \Nscr\setminus\Lscr(S_j)$ is not bad for $S_j$.
At this point one uses the particular choice of the decomposition of $x^*$ into incidence vectors of spanning trees.
For every edge $e$ that is not bad we have $c^j(e) =c(e)$.
Hence
\begin{eqnarray*}
%\min \{c^j(J) : J\text{ a } T_j\text{-join} \}
%&\le& 
c^j(y^j_{\beta}) 
&=& \beta \, c^j(x^*) \ +\ \alpha \, c(S_j)  + \sum_{C\in\Lscr(S_j)} \beta(2-x^*(C)) \, c(S_j \cap C) \\
&& + \sum_{C\in\Nscr\setminus\Lscr(S_j)} \max\left\{0,\, \beta(2-x^*(C)) - \alpha \big. \right\} c(v^C).
\end{eqnarray*}
\bigskip
Moreover, Seb\H{o} and van Zuylen \cite{SebvZ16} showed:
\bigskip
\begin{lemma} 
%For every spanning tree $S\in \{S_1, \dots, S_r\}$
$$c^j(x^*) \ \le \ c(x^*) + \sum_{C\in\Lscr(S_j)} 2(x^*(C)-1) c(S_j\cap C).$$
\end{lemma}

Therefore, the cost of the tour that results from the tree $S_j$ is at most
\begin{equation}\label{eq:cost_single_tour}
\begin{aligned}
 &c(F_{j}) + \min \{c^{j}(J) : J\text{ a } T_{j}\text{-join} \} \\
 \le \ &c(S_{j}) - \!\sum_{C\in\Lscr(S_{j})} c(S_{j}\cap C)  + c^{j}(y^j_{\beta}) \\
 \le \ &(1+\alpha)c(S_j) + \beta c(x^*) + \sum_{C\in\Lscr(S_{j})}\bigl(2\beta(x^*(C)-1)- 1 + \beta(2-x^*(C)) \bigr) \, c(S_{j}\cap C)\\
 & + \sum_{C\in\Nscr\setminus\Lscr(S_j)} \max\left\{0,\, \beta (2-x^*(C)) - \alpha \big. \right\} c(v^C) \\
 =\  &(1+\alpha)c(S_j) + \beta c(x^*) - \sum_{C\in\Lscr(S_{j})}\bigl(\alpha+ \beta (2-x^*(C)) \bigr) \, c(S_{j}\cap C)\\
 & + \sum_{C\in\Nscr\setminus\Lscr(S_j)} \max\left\{0,\, \beta(2-x^*(C)) - \alpha \big. \right\} c(v^C),
\end{aligned}
\end{equation}
since $\alpha = 1 - 2\beta$.

\section{Average cost}

It will be useful to index the trees by a continuum and define $S_{\sigma}$ for all $0<\sigma\le 1$, where
$S_{\sigma}=S_j$ if $\sum_{i=1}^{j-1} p_i <\sigma \le \sum_{i=1}^j p_i$.

Let $h\colon [0,1]\to[0,1]$ be an integrable function to be chosen later.
The weight of the tour resulting from $S_{\sigma}$ will be proportional to $1+h(\sigma)$.
Also $\alpha$ and $\beta$ depend on $\sigma$, namely as follows:
$$\alpha_{\sigma} = \frac{1-h(\sigma)}{1+h(\sigma)}   \qquad \text{ and } \qquad \beta_{\sigma} = \frac{h(\sigma)}{1+h(\sigma)}.$$
Note that indeed $0\le \beta_{\sigma}\le\frac{1}{2}$ and $\alpha_{\sigma}+2\beta_{\sigma}=1$ for all $\sigma$.

Moreover, we set
\[ v^C \ \coloneqq \ \frac{1}{\int_0^z \bigl( 1 - h(\sigma) + z h(\sigma) \bigr) \,\text{d}\sigma} \,
\int_{0}^{z} \bigl( 1 - h(\sigma) + z h(\sigma) \bigr)  \cdot \chi^{S_\sigma\cap C}  \,\text{d}\sigma, \]
where we abbreviated $z\coloneqq 2-x^*(C)$.
Then indeed $v^C(C) = 1$ for all $C\in\Nscr$, and $v^C(e) = 0$ unless $e$ is lonely at $C$.

We will now show under which condition the last two terms in \eqref{eq:cost_single_tour} vanish:

\begin{lemma}\label{lemma:lonely_edge_usage}
Suppose
\begin{equation}
\label{eq:requirementforh}
\int_z^1 \max \bigl\{0, h(\sigma)-1+zh(\sigma) \bigr\} \,\text{d}\sigma + \int_0^z \bigl( h(\sigma)-1-zh(\sigma) \bigr) \,\text{d}\sigma \ \le \ 0
\end{equation}
for all $z\in[0,1]$. 
Then
\begin{equation}
\label{eq:vanish}
\begin{aligned}
\int_0^1  (1+h(\sigma)) \Biggl( \,
& - \sum_{C\in \Lscr(S_{\sigma})} \!
 \bigl( 
 \alpha_{\sigma} + \beta_{\sigma} (2-x^*(C)) \bigr)\, c(S_{\sigma}\cap C) \\
+&\!\! \sum_{C\in\Nscr\setminus\Lscr(S_j)} \!\!
\max\left\{0,\, \beta_{\sigma} (2-x^*(C)) - \alpha_{\sigma} \big. \right\} c(v^C) 
\Biggr) \,\textnormal{d}\sigma 
\end{aligned}
\end{equation}
is nonpositive.
\end{lemma}
\prove 
Again writing $z\coloneqq 2-x^*(C)$, using
$$ (1+h(\sigma))  \bigl( \alpha_{\sigma} + \beta_{\sigma} (2- x^*(C)) \bigr) \ = \ 
 1 - h(\sigma) + z h(\sigma) $$
and
$$ (1+h(\sigma)) \max\left\{0,\, \beta_{\sigma} (2-x^*(C)) - \alpha_{\sigma} \big. \right\} 
\ = \  \max\left\{0,\ h(\sigma) - 1 + z h(\sigma) \big. \right\},$$
and changing the order of summation, 
we can rewrite \eqref{eq:vanish} as
\begin{align*}
- \sum_{C\in\Nscr}  \int_0^z  \bigl( 1 - h(\sigma) + z h(\sigma) \bigr)\, c(S_{\sigma}\cap C) \,\textnormal{d}\sigma
+  \sum_{C\in\Nscr} \int_z^1 \max\left\{0,\,  h(\sigma) - 1 + z h(\sigma) \big. \right\} c(v^C) \,\textnormal{d}\sigma.
\end{align*}

Hence (plugging in the definition of $v^C$) and using $1-h(\sigma) +z \cdot h(\sigma)>0$, it suffices to show that, for every $z\in(0,1]$, 

$$-1 + \frac{1}{\int_0^z  \bigl( 1 - h(\sigma) + z h(\sigma) \bigr) 
\,\text{d}\sigma} \int_z^1 \max\left\{0,\, h(\sigma) - 1 + z h(\sigma) \big. \right\} d\sigma\ \le \ 0. $$
which follows directly from \eqref{eq:requirementforh}.
\endproof

\begin{lemma}
Let $h\colon [0,1]\to[0,1]$ be an integrable function with \eqref{eq:requirementforh} for all $z\in[0,1]$.
Then the best-of-many Christofides algorithm with lonely edge deletion computes a solution of cost 
at most $\rho^* c(x^*)$, where
$$\rho^* \ = \ 1 + \frac{1}{1 + \int_0^1 h(\sigma) \,\textnormal{d}\sigma}.$$
\end{lemma}

\prove
Combining \eqref{eq:cost_single_tour} and Lemma \ref{lemma:lonely_edge_usage},
we get the following upper bound on the total cost of the best-of-many Christofides algorithm with lonely edge deletion:
 \begin{align*}
&  \sfrac{1}{\int_0^1 (1+h(\sigma)) \,\text{d}\sigma}  \int_0^1 (1+h(\sigma)) 
\left( \Big. \beta_{\sigma} c(x^*) +  (1+\alpha_{\sigma}) c(S_{\sigma})   \right) \,\text{d}\sigma \\
= \ & \sfrac{1}{\int_0^1 (1+h(\sigma)) \,\text{d}\sigma}  \int_0^1  
 \left( \Big. h(\sigma) c(x^*) + 2 c(S_{\sigma}) \right) \,\text{d}\sigma \\
= \ &  \sfrac{1}{\int_0^1 (1+h(\sigma)) \,\text{d}\sigma} \, \left( \int_0^1 h(\sigma) \,\text{d}\sigma + 2 \right) c(x^*) \\
= \ &  \sfrac{1}{\int_0^1 (1+h(\sigma)) \,\text{d}\sigma} \, \left( \int_0^1 (1+h(\sigma)) \,\text{d}\sigma + 1  \right) c(x^*) \\
= \ & \left( 1 +  \sfrac{1}{1 + \int_0^1 h(\sigma) \,\text{d}\sigma} \right) c(x^*)
\end{align*}
\endproof

Now we can prove the main result:

\begin{theorem}\label{theorem:upper_bound_rho}
Let 
$$\rho^* \ \coloneqq \ 1 + \frac{1}{1 + 4 \ln(\frac{5}{4})}.$$ 
Then the best-of-many Christofides algorithm with lonely edge deletion computes a solution of cost 
at most $\rho^* c(x^*)$. 
\end{theorem}

\prove
We set $h(\sigma)=\frac{4}{4+\sigma}$ for $0\le\sigma\le 1$.
Then $\int_0^1 \sfrac{4}{4+\sigma} \,\text{d}\sigma =  4\ln(\frac{5}{4})$.
We need to check  \eqref{eq:requirementforh}.

Note that $h(\sigma)-1+z h(\sigma) >0$ if and only if $\sfrac{4}{4+\sigma} = h(\sigma)> \frac{1}{1+z}$, i.e., $\sigma<4z$.
Hence to prove \eqref{eq:requirementforh} it suffices to show
$$\int_z^{4z} \bigl( h(\sigma)-1+zh(\sigma) \bigr) \,\text{d}\sigma + \int_0^z \bigl( h(\sigma)-1-zh(\sigma) \bigr) \,\text{d}\sigma \ \le \ 0$$
The left-hand side is
$$4(1+z) ( \ln (4z+4) - \ln(z+4) ) + 4(1-z) ( \ln (z+4) - \ln(4) ) -4z,$$
so (dividing by 4) we need to check
$$(1+z) \ln \sfrac{4z+4}{z+4} + (1-z) \ln \sfrac{z+4}{4} - z \ \le \ 0.$$
This is true for $z=0$, moreover the derivative of the left-hand side is
$$\ln \sfrac{16(z+1)}{(z+4)^2} -\sfrac{2z}{z+4}.$$
Using $\ln x \le x-1$ for all $x>0$ this is at most
$$ \sfrac{16(z+1)}{(z+4)^2} - 1 -\sfrac{2z}{z+4} 
\ = \  \sfrac{16(z+1)-(z+4)^2 -2z(z+4)}{(z+4)^2} 
\ = \  \sfrac{-3z^2}{(z+4)^2} 
\ \le \ 0.$$
\endproof

Theorem \ref{theorem:upper_bound_rho} immediately implies that the integrality ratio is at most $\rho^*$.
Note that $\rho^* < 1.5284$.
We see that \eqref{eq:requirementforh} is tight only for $z=0$ with our choice of $h$. A better choice would lead to a better 
upper bound on the integrality ratio. 
However, we do not know how to find the best $h$.
Numerical computations indicate that the best value that can be obtained in this way is approximately $1.5273$.

\subsubsection*{Acknowledgment} 
We thank the anonymous referees for their useful remarks that helped to improve the presentation. 
Moreover, one referee suggested a Python-Gurobi script to compute
the best bound on the integrality ratio that can be obtained by our approach numerically.

\end{document}